\documentclass[12pt]{iopart}
\usepackage{iopams}  
\usepackage{setstack}  
\usepackage{epsfig}
\usepackage{hypernat}
\usepackage{color}
\usepackage{ifpdf}
\usepackage{caption}
\usepackage[english]{babel}
\newif\ifcomment
\newif\ifprint
\newif\ifshort
\shorttrue

\ifpdf
\ifprint
 \usepackage[plainpages=false,pdfpagelabels,
             linkbordercolor={0 0 .8},%
             citebordercolor={.8 0 0},%
             urlbordercolor={.8 0 .8},%
             raiselinks=true,%
             pdfborder={0 0 1 [3]}]{hyperref}
\else
 \usepackage[pdftex,backref,pagebackref,colorlinks,
             plainpages=false,pdfpagelabels,
             linkcolor=blue,citecolor=blue,anchorcolor=blue,
             pagecolor=blue,urlcolor=red,a4paper]{hyperref}
\fi
\pdfoutput=1        
\pdfcatalog{/PageMode(/UseOutlines)} 
\else 
\ifprint
 \usepackage[plainpages=false,pdfpagelabels,
             linkbordercolor={0 0 .8},%
             citebordercolor={.8 0 0},%
             urlbordercolor={.8 0 .8},%
             raiselinks=true,%
             pdfborder={0 0 1 [3]}]{hyperref}
\else
 \usepackage[dvips,backref,pagebackref,colorlinks,
             plainpages=false,pdfpagelabels,
             linkcolor=blue,citecolor=blue,anchorcolor=blue,
             pagecolor=blue,urlcolor=red,a4paper]{hyperref}
\fi
\hypersetup{
  pdfauthor   = {Constantin Loizides},
  pdftitle    = {},
  pdfcreator  = {},
  pdfproducer = {},
  pdfsubject  = {},
  pdfkeywords = {}
 }
\fi

\graphicspath{{./pics/}}

\def\PHOBOS{P\kern-.34em \lower.4ex\hbox{H}\kern-.12em \lower.6ex\hbox{O}\kern-.08em \lower-.18ex\hbox{B}\kern-.12em \lower-.45ex\hbox{O}\kern-.12em \lower.12ex\hbox{S}\ }
\def\eps{\epsilon}
\def\mycol{red}
\newcommand {\snn}       {\ensuremath{\sqrt{s_{\rm NN}}}}

\newcommand {\av}[1]     {\ensuremath{\left< #1 \right>}}
\newcommand {\abs}[1]    {\ensuremath{\left| #1 \right|}}
\newcommand {\txt}[1]    {\ensuremath{\mathrm{#1}}}
\newcommand {\hrefurl}[1]{\href{#1}{\url{#1}}}
\newcommand {\arxiv}[1]  {\href{http://www.arxiv.org/#1}{\mbox{Preprint #1}}}
\newcommand {\Ref}[1]    {Ref.~\cite{#1}}

\newcommand {\Eq}[1]     {Equation~(\ref{#1})}

\newcommand {\Fig}[1]    {Fig.~\ref{#1}}

\begin{document}

\title[Elliptic flow fluctuations in $\sqrt{s_{NN}}=200$~GeV Au+Au collisions at RHIC]
{Elliptic flow fluctuations in $\mathbf{\sqrt{s_{NN}}=200}$~GeV Au+Au collisions at RHIC}

\ifprint
\else
\author{Constantin Loizides for the PHOBOS collaboration}
\address{
\vspace{2mm}
%
%
{\footnotesize
%
%
B~Alver$^4$,
B~B~Back$^1$,
M~D~Baker$^2$,
M~Ballintijn$^4$,
D~S~Barton$^2$,
R~R~Betts$^6$,
A~A~Bickley$^7$,
R~Bindel$^7$,
W~Busza$^4$,
A~Carroll$^2$,
Z~Chai$^2$,
V~Chetluru$^6$,
M~P~Decowski$^4$,
E~Garc\'{\i}a$^6$,
N~George$^2$,
T~Gburek$^3$,
K~Gulbrandsen$^4$,
C~Halliwell$^6$,
J~Hamblen$^8$,
I~Harnarine$^6$,
M~Hauer$^2$,
C~Henderson$^4$,
D~J~Hofman$^6$,
R~S~Hollis$^6$,
R~Ho\l y\'{n}ski$^3$,
B~Holzman$^2$,
A~Iordanova$^6$,
E~Johnson$^8$,
J~L~Kane$^4$,
N~Khan$^8$,
P~Kulinich$^4$,
C~M~Kuo$^5$,
W~Li$^4$,
W~T~Lin$^5$,
C~Loizides$^4$,
S~Manly$^8$,
A~C~Mignerey$^7$,
R~Nouicer$^2$,
A~Olszewski$^3$,
R~Pak$^2$,
C~Reed$^4$,
E~Richardson$^7$,
C~Roland$^4$,
G~Roland$^4$,
J~Sagerer$^6$,
H~Seals$^2$,
I~Sedykh$^2$,
C~E~Smith$^6$,
M~A~Stankiewicz$^2$,
P~Steinberg$^2$,
G~S~F~Stephans$^4$,
A~Sukhanov$^2$,
A~Szostak$^2$,
M~B~Tonjes$^7$,
A~Trzupek$^3$,
C~Vale$^4$,
G~J~van~Nieuwenhuizen$^4$,
S~S~Vaurynovich$^4$,
R~Verdier$^4$,
G~I~Veres$^4$,
P~Walters$^8$,
E~Wenger$^4$,
D~Willhelm$^7$,
F~L~H~Wolfs$^8$,
B~Wosiek$^3$,
K~Wo\'{z}niak$^3$,
S~Wyngaardt$^2$,
B~Wys\l ouch$^4$\\
}
\vspace{2mm}
%
%
%
%
{\scriptsize
$^1$~Argonne National Laboratory, Argonne, IL 60439-4843, USA\\
$^2$~Brookhaven National Laboratory, Upton, NY 11973-5000, USA\\
$^3$~Institute of Nuclear Physics PAN, Krak\'{o}w, Poland\\
$^4$~Massachusetts Institute of Technology, Cambridge, MA 02139-4307, USA\\
$^5$~National Central University, Chung-Li, Taiwan\\
$^6$~University of Illinois at Chicago, Chicago, IL 60607-7059, USA\\
$^7$~University of Maryland, College Park, MD 20742, USA\\
$^8$~University of Rochester, Rochester, NY 14627, USA\\
}
}
\color{black}
\fi

\begin{abstract}
We present first results on event-by-event elliptic flow fluctuations in 
Au+Au collisions at $\sqrt{s_{\rm NN}}=200$ GeV obtained with the PHOBOS 
detector. Over the measured range in centrality, large relative fluctuations 
of \mbox{$40$--$50$\%} are found. 
The elliptic flow fluctuations are well described as being proportional
to fluctuations in the shape of the initial collision region, as estimated 
event-by-event with the participant eccentricity using Glauber 
Monte Carlo. 
\end{abstract}

\ifprint
\else
\submitto{\JPG}
\color{black}
\fi

\section{Introduction}
Elliptic flow~($v_2$) 
is one of the key observables in the understanding of the dynamics 
of heavy ion collisions.
In the collision of two nuclei with finite impact parameter, 
the overlap region in the plane transverse to the beam direction
is spatially asymmetric. 
Observation of an azimuthal anisotropy in the momentum distributions of the
detected particles relative to the reaction plane 
is direct evidence 
of interactions between the initially produced particles~\cite{poskanzer98}.
These interactions must occur at relatively early times, since the expansion 
of the source ---even if uniform--- 
would wash out the effect of the initial asymmetry.
Quantitatively, the connection between the initial and final-state anisotropy 
is exploited by hydrodynamical models that relate a given initial source shape 
to the distribution of produced particles.
In such calculations, one typically uses smooth, event-averaged, initial conditions.
However, event-by-event fluctuations in the shape of the initial interaction region 
must not be neglected.
\ifshort
As a means to quantify the effect of initial-state eccentricity fluctuations, 
we have introduced the ``participant eccentricity'', $\eps_{\rm part}$.
It unifies the eccentricity-scaled elliptic flow across Cu+Cu and Au+Au 
collisions~\cite{ManlyQM05,PhobosFlowPRL3}. 
It 
\else
\color{\mycol}
To quantify the effect of initial-state eccentricity fluctuations, 
we have introduced the ``participant eccentricity'',
$ \eps_{\rm part} = \frac{\sqrt{(\sigma_{y}^2-\sigma_{x}^2)^2+4\sigma_{xy}^2}}
                         {\sigma_{y}^2+\sigma_{x}^2}$, 
where $\sigma^2_{x}$,  $\sigma^2_{y}$ and 
$\sigma_{xy}=\langle xy\rangle - \langle x \rangle\langle y\rangle$ 
are the per-event \mbox{(co-)variances} of the participant nucleon position 
distributions in the transverse plane.
It unifies the eccentricity-scaled flow, 
$v_2/\langle \eps_{part} \rangle$, for Cu+Cu and Au+Au 
collisions, as a function of particle area density~(or centrality). 
Its magnitude and shape as a function of centrality are rather robust to 
variations of the Glauber parameters~\cite{ManlyQM05,PhobosFlowPRL3}.
The participant eccentricity
\color{black}
\fi
accounts for nucleon-position fluctuations in the participating nucleon
distributions by calculating the eccentricity, event-by-event, 
with respect to the principal axes of the overlap ellipse in a 
Monte Carlo Glauber~(MCG) simulation. 
In a hydrodynamical scenario, such fluctuations in the shape of the initial 
collision region would lead naturally to corresponding fluctuations in the 
elliptic flow signal. 
\ifshort
\else
\color{\mycol}
To estimate their magnitude, we assume that $v_2\propto\eps$, derived from hydrodynamical 
calculations, holds event-by-event. This leads to
${\sigma_{v_2}}/{\langle v_2 \rangle} = {\sigma_{\eps}}/{\langle \eps \rangle}$,
where $\sigma_{v_2}$~($\sigma_{\eps}$) is the standard deviation of the 
event-by-event distribution $v_2$~($\eps$).
Neglecting all other sources of elliptic flow fluctuations,
our participant eccentricity MCG simulation predicts relative 
fluctuations~($\sigma_{v_2}/\langle v_2 \rangle$) of \mbox{$35$--$50$\%}
in Au+Au collisions at $\snn=200$~GeV.
For all but the most central \mbox{$0$--$6$\%} collisions, this is more than 
a factor of 2 larger than estimations based on merely $N_{\rm part}$ driven fluctuations, 
and of similar size as expected from statistical fluctuations.
\color{black}
\fi
Note that the measurement of $v_2$ fluctuations as a probe of early stage dynamics 
of heavy ion collisions has been suggested earlier~\cite{Shuryak}, although without
considering the possibility of initial-state eccentricity fluctuations.
In order to experimentally determine elliptic flow fluctuations, 
we have developed an analysis technique for the extraction of dynamical flow 
fluctuations~\cite{Alver06}.
Our method, summarized below, relies crucially on the large pseudo-rapidity coverage 
of the PHOBOS multiplicity array, which detects a very large fraction of the produced 
particles~\cite{WhitePaper}. The results as a function of centrality in $\snn=200$~GeV 
Au+Au collisions at mid-rapidity are for the first time presented in these conference 
proceedings.

\section{Elliptic flow fluctuations analysis technique and results}
\ifshort
In order to disentangle known~(mostly statistical) from unknown~(dynamical) 
contri-butions to the measured flow fluctuations, a detailed knowledge of the detector 
response is required. The response function, $K(v_{2}^{\txt{obs}},v_{2}, n)$, is defined 
as the distribution of the event-by-event observed elliptic flow,~$v_{2}^{\txt{obs}}$, 
for events with constant input flow value,~$v_2$, and multiplicity,~$n$. It is designed
to account for detector deficiencies, as well as for multiplicity and finite-number 
fluctuations. 
\else
\color{\mycol}
Our analysis method is designed to disentangle known~(mostly statistical) 
from unknown~(dynamical) contributions to the measured flow fluctuations. 
Typically, one assumes that both contributions add in quadrature, according to
\mbox{$\sigma_{v_{2},\txt{obs}}^2 = \sigma_{v_{2},\txt{dyn}}^2 + \sigma_{v_{2},\txt{stat}}^2$}.
This relation holds, if the average of the measurement, $\langle
v_{2}^{\txt{obs}}\rangle$, agrees with the true average in the data, $\langle
v_{2}\rangle$, and if the resolution of the method is independent of
the true value. Since neither of both conditions are fully satisfied in our
event-by-event measurement of $v_2$, a more detailed knowledge of the detector
response is required. 
The resolution depends on the magnitude of the flow signal and event 
multiplicity. Thus, we define the response function, $K(v_{2}^{\txt{obs}},v_{2}, n)$, 
as the distribution of the event-by-event observed elliptic flow,~$v_{2}^{\txt{obs}}$, 
for events with constant input flow value,~$v_2$, and multiplicity,~$n$. As explained
below, it is not only designed to account for detector deficiencies, but also for
multiplicity and finite-number fluctuations. 
\color{black}
\fi
Let $f(v_2)$ be the true 
$v_2$ distribution for a set of events in a given centrality class, 
and $g(v_{2}^{\txt{obs}})$ the corresponding distribution of the observed~$v_{2}$, 
$v_{2}^{\txt{obs}}$.
The true and observed distributions are related by
\begin{equation}
 g(v_{2}^{obs}) = \int K(v_{2}^{\txt{obs}},v_2,n) \, f(v_2) 
 \, N(n) \, \txt{d}v_2  \, \txt{d}n\,,
 \label{eqkernel}
\end{equation}
where $N(n)$ is the multiplicity distribution of the events in the given set of 
events. Our analysis method consists of three steps:
\ifshort
\else
\begin{enumerate}
\color{\mycol}
\item Measurement of the observed $v_2^{\txt{obs}}$ distribution, $g(v_{2}^{\txt{obs}})$, 
for a set of events by measuring $v_{2}^{\txt{obs}}$ event-by-event.
\item Construction of the kernel, $K(v_{2}^{\txt{obs}},v_2,n)$, by studying the
detector response for Monte Carlo events with constant~(known) input value of $v_2$ 
and multiplicity $n$. 
\item Determination of the mean and variance of the true $v_2$ distribution, $f(v_2)$, 
by solving \Eq{eqkernel} with a functional ansatz for $f(v_2)$.
\end{enumerate}
\color{black}
\fi

i) Measurement of the $v_2^{\txt{obs}}$ distribution, $g(v_{2}^{\txt{obs}})$:
The event-by-event measurement utilizes a two-parameter maximum likelihood fit 
to the obtained hit information in an event. 
\ifshort
\else
\color{\mycol}
The fit makes use of all hits from the multiplicity array for the measurement of a single value, 
$v_{2}^{\txt{obs}}$, while it efficiently corrects for non-uniformities in the acceptance of the 
participating sub-detector systems. 
\color{black}
\fi
The probability distribution function~(PDF) of a particle to be emitted in the 
direction~($\eta,\phi$) for an event with mid-rapidity flow amplitude 
$v_2\equiv v_2(0)$ and reaction plane angle $\phi_0$ is 
defined as \mbox{$P(\eta,\phi;v_2,\phi_{0}) = 
s^{-1}\,[1+2v_{2}(\eta)\cos(2\phi-2\phi_0)]$}.
The pseudorapidity dependence, $v_2(\eta)$, is parametrized with a triangular shape, 
\mbox{$v^{\rm tri}_2(\eta)=v_2\,(1-\frac{\abs{\eta}}{6})$},
or alternatively with a trapezoidal shape, 
\mbox{$v^{\rm trap}_2(\eta) = \left\{^{v_2\,\txt{if}\,\abs{\eta}<2\,}_{\frac{3}{2}\,
v^{\rm tri}_2(\eta)\,\txt{else}}\right.$}. 
Both parametrizations 
yield good descriptions of the previously measured~(mean) $v_2(\eta)$ 
shapes~\cite{PhobosFlowPRC}. 
The normalization~($s^{-1}$) is constructed such that the PDF folded with 
the PHOBOS acceptance~($A_{\rm cc}$) yields the same value for different sets of 
parameters~($v_2,\phi_0$),
$s(v_2,\phi_{0};\eta) = \int_{\eta-\Delta\eta/2}^{\eta+\Delta\eta/2} A_{\rm cc}(\eta',\phi')\,
[1+2 v_2(\eta')\cos(2\phi'-2\phi_{0})] \, \txt{d}\phi'\,\txt{d}\eta'$, in practice
constructed in small bins of $\Delta\eta$.
The desired value $v^{\txt{obs}}_{2}$ is given by the 
pair~($v^{\txt{obs}}_{2},\phi^{\txt{obs}}_0$)
that is ``most likely'' for the given event, found by maximizing 
$\prod_{i=1}^{n_{\rm hits}} P(\eta_i,\phi_i;v_2,\phi_0)$ as a function of 
$v_2$ and $\phi_0$ for all hits in the event.

ii) Construction of the kernel, $K(v_{2}^{\txt{obs}},v_2,n)$: 
The response function $K(v_2^{\txt{obs}},v_2,n)$ is determined by running
the event-by-event analysis, as described above, on modified HIJING events with
known values of $v_2$ and $n$. 
Flow of constant magnitude~($v_2$) with a flat reaction plane distribution~($\phi_0$)
is introduced into the generated HIJING events. This is achieved by redistributing 
the resulting particles in each event in $\phi$ according to 
$1+2v_{2}(\eta)\cos(2\phi-2\phi_{0})$, using their generated $\eta$ positions. 
The modified HIJING events are fed through GEANT \ifcomment~\cite{GEANT}\fi to 
obtain the PHOBOS detector response. 
For each parametrization of $v_2(\eta)$, triangular and trapezoidal, used in the 
event-by-event measurement, a corresponding kernel, $K^{\rm tri}$ and $K^{\rm trap}$, 
is constructed.
To obtain the kernel in bins of $v_2$ and $n$ with enough precision would require 
on the order of 100 million MC events. Instead the kernel can be parametrized, 
allowing us to use about 1.5\% of that statistics to reach the required precision. 
The response of a perfect detector can be determined as a function of event multiplicity
as described in \Ref{Ollitrault1992}~(see Eq.~(A13) with $\alpha \rightarrow v_2^{\txt{obs}}$, 
$\overline{\alpha} \rightarrow v_2$ and $M \rightarrow n$).
In practice, however, it turns out that $v_2$ is suppressed, with the suppression dependent 
on $n$, and that the resolution~($\sigma$) has a constant background contribution. 
With $v_{2}^{\txt{sup}}=(A\,n+B)\,v_{2}$ and $\sigma = C/\sqrt{n} + D$, this leads to
\begin{equation}
 K(v_{2}^{\txt{obs}},v_{2},n) = \frac{v_2^{\txt{obs}}}{\sigma^2} \, 
     \exp  \left(-\frac{\left(v_2^{\txt{obs}}\right)^2+ \left(v_2^{\txt{sup}}\right)^2}{2\sigma^2}\right)
     I_{0} \left(-\frac{v_{2}^{\txt{obs}}v_2^{\txt{sup}}}{\sigma^2}\right)
\end{equation}
where $I_0$ is a modified Bessel function of the first kind. 
The four unknown parameters~($A,B,C,D$) are obtained using the modified HIJING samples.

iii) Determination of the mean and variance of the true $v_2$ distribution, $f(v_2)$:
To extract the fluctuations, we assume a Gaussian distribution for $f(v_2)$, with two 
parameters, $\langle v_{2}\rangle$ and $\sigma_{v_2}$. For given values of $\langle v_{2}\rangle$
and $\sigma_{v_2}$, it is possible to take the integral in \Eq{eqkernel} to obtain the
expected distribution, \mbox{$g_{\txt{exp}}(v_2^{\txt{obs}};\langle
v_{2}\rangle,\sigma_{v_2})$}. By comparing the expected and observed distributions,
the values for $\langle v_2 \rangle$ and $\sigma_{v_2}$ are found by a maximum likelihood
fit. 

\begin{figure}[t]
\captionsetup[figure]{margin=0.1cm,font=small}

\begin{minipage}[t]{0.5\textwidth}
 \centerline{
  \includegraphics[width=\textwidth]{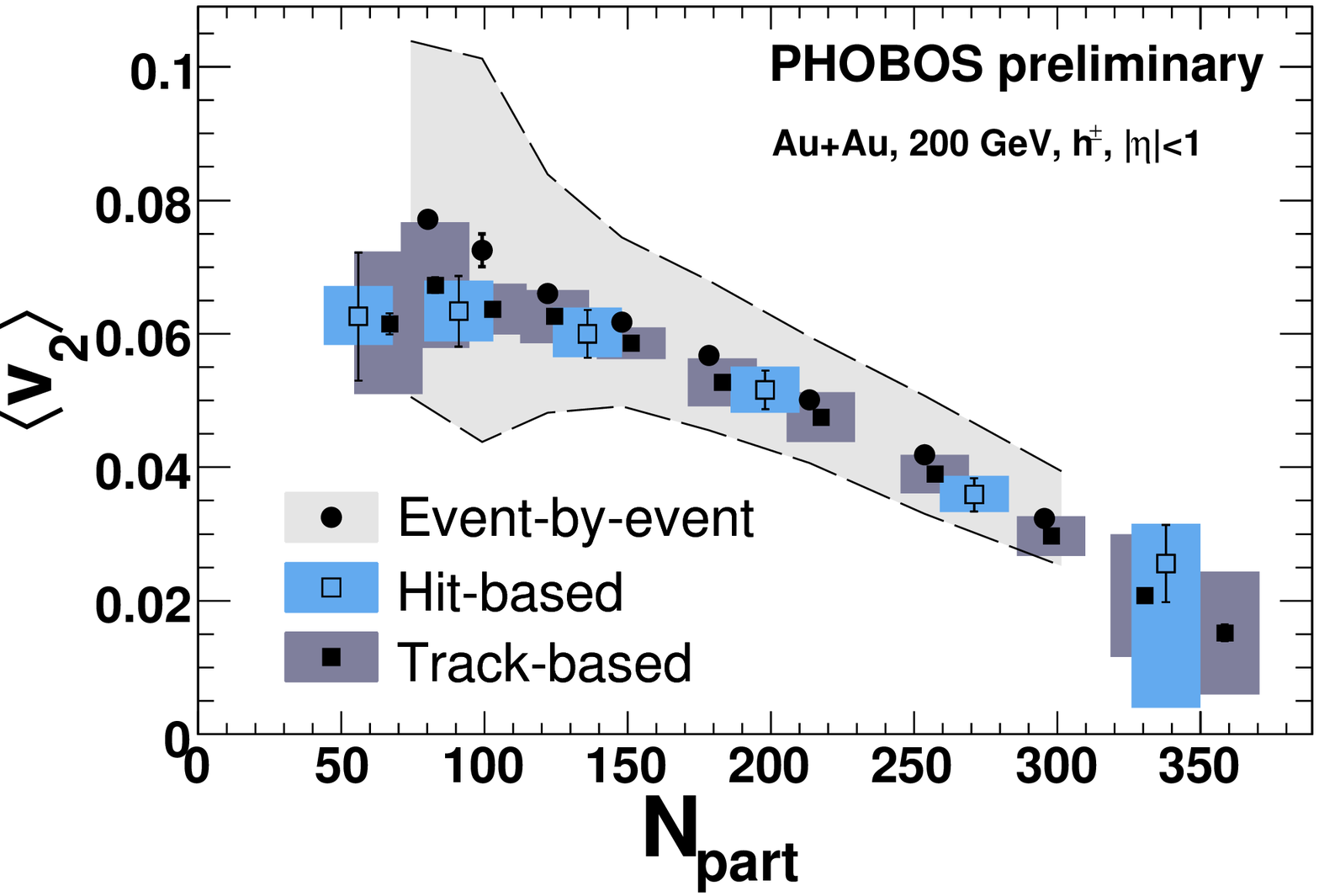}
 }
 \vspace{-0.2cm}
 \caption{\label{fig1} Mean flow, $\av{v_2}$, 
          as a function of centrality, 
          for $\snn=200$~GeV collisions at mid-rapidity, measured
          by the event-by-event analysis, compared to the published 
          results obtained with the event-averaged, sub-event based 
          technique using hits or tracks~\cite{PhobosFlowPRC}.}
\end{minipage}
\hspace{\fill}
\begin{minipage}[t]{0.5\textwidth}
 \centerline{
  \includegraphics[width=\textwidth]{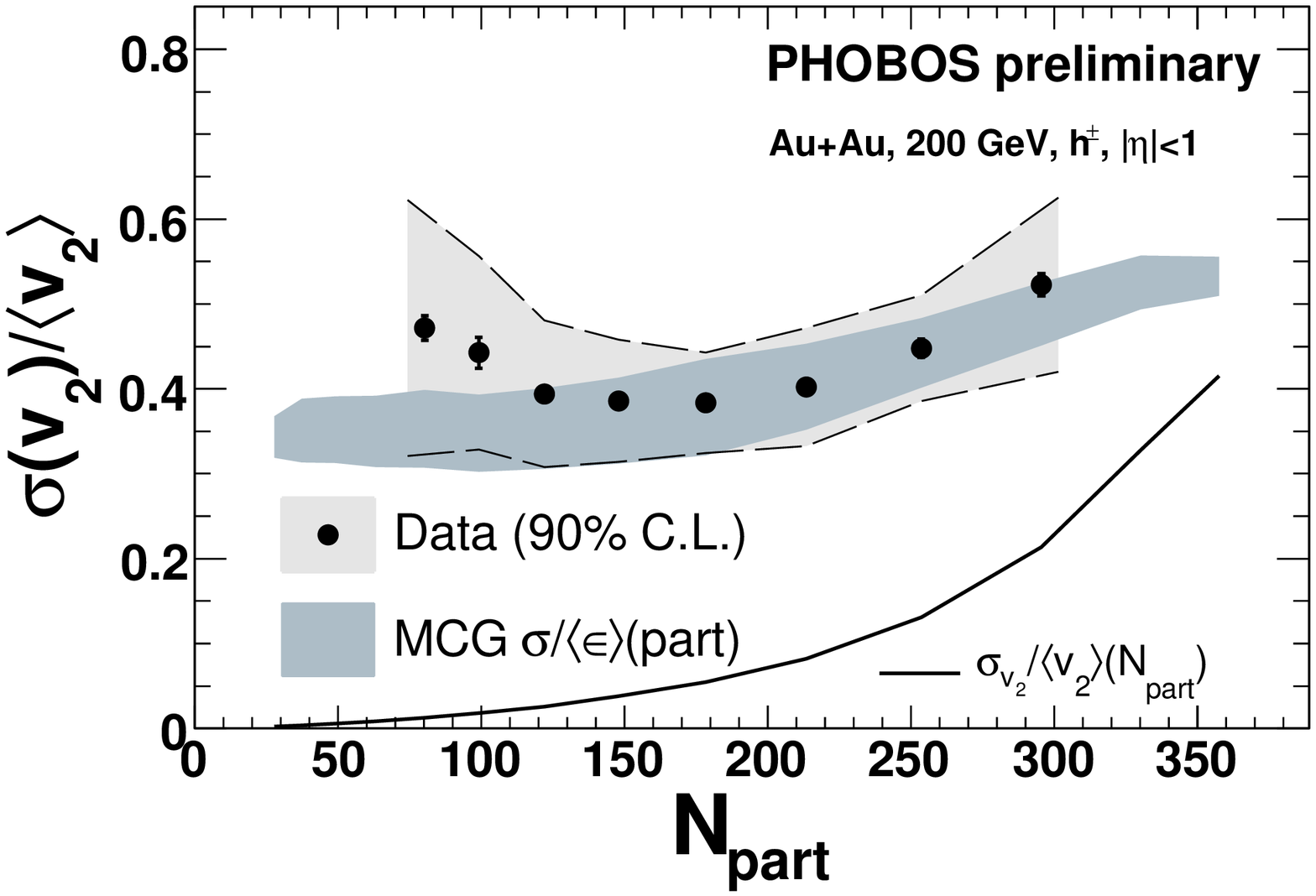}
 }
 \vspace{-0.2cm}
 \caption{\label{fig2} Relative flow fluctuations,~$\sigma_{v_2}/\av{v_2}$,
          as a function of centrality, 
          for \mbox{$\snn=200$} GeV collisions at mid-rapidity, 
          compared to the prediction, $\sigma_{\eps}/\av{\eps}$,
          from the participant eccentricity, and to an estimate of
          $N_{\rm part}$-induced fluctuations 
          using a fit of $\av{v_2}(N_{\rm part})$.
}
\end{minipage}
\end{figure}

The analysis chain is applied to the $\snn=200$~GeV Au+Au data set from Run~4
in bins of centrality. The results~($\langle v_{2}\rangle$ and $\sigma_{v_2}$) are obtained 
separately for triangular and trapezoidal $v_2(\eta)$ shape and averaged over bins of
collision vertex position. 
The systematic errors~(all sources added in quadrature) are estimated by including 
variations from different vertex and $\phi_0^{\txt{obs}}$ bins and changes introduced 
by the triangular and trapezoidal $v_2(\eta)$ shapes. Since the functional form of the 
true distribution is unknown, differences arising from a flat rather than a Gaussian 
ansatz for $f(v_2)$ are included. Furthermore, we have performed extensive studies of the 
analysis response to MC samples, prepared to match the observed multiplicity distribution 
and $\langle v_2\rangle$, in bins of centrality and known input values of 
fluctuations~(including zero). The discrepancies of the obtained to the input fluctuations 
evaluated per centrality bin 
are an additional source of the systematic error.
Their contribution becomes increasingly important for low values of $v_2$, and
therefore the most central collision bin (\mbox{$0$--$6$\%}) was excluded from the analysis.
\Fig{fig1} shows the mean flow results obtained with the event-by-event 
analysis projected to mid-rapidity by 
$\av{v_2}=0.5\left(\frac{11}{12} \av{v_2^{\rm tri}} + \av{v_2^{\rm trap}}\right)$,
which ---within the systematic errors--- agree very well with the published results 
of the event-averaged sub-event based technique using hits or tracks~\cite{PhobosFlowPRC}. 
This constitutes an important, completely independent verification of the event-by-event 
analysis. \Fig{fig2} presents the relative flow fluctuations,~$\sigma_{v_2}/\av{v_2}$.
In the ratio, most of the aforementioned systematic errors scale out. We find large 
relative fluctuations of about \mbox{$40$--$50$\%}, with little centrality dependence. 
The data are well described by the fluctuations in the participants
eccentricity, $\sigma_{\eps}/\av{\eps}$, obtained in MCG simulations. 
These are shown as the shaded band in \Fig{fig2}, corresponding to 90\%
confidence level boundaries obtained by varying the Glauber parameters within
reasonable limits~(see \Ref{PhobosFlowPRL3}). 
The contribution from $N_{\rm part}$ fluctuations is estimated using a fit to
the $\av{v_2}$ data folded with the known $N_{\rm part}$ distributions from 
the PHOBOS centrality trigger studies. Over the range of measured
centrality, their contribution~(taken in quadrature) is 
small, less than 15\%.

\ifprint
\else
\newpage
\section{Conclusions}
\fi
First results on event-by-event elliptic flow fluctuations in 
Au+Au collisions at \mbox{$\sqrt{s_{\rm NN}}=200$}~GeV measured
with the PHOBOS detector are presented.
\ifshort
\else
\color{\mycol}
Our analysis procedure for extracting dynamical fluctuations consists
of three major steps: i)~The elliptic flow~($v_2$) is determined 
event-by-event by a maximum likelihood fit using its known pseudo-rapidity
dependence. ii)~For a given centrality bin, the detector response function 
is calculated using detailed Monte Carlo simulations with varying input 
$v_2$ values. iii)~The dynamical fluctuations are determined by solving the 
response function with a functional ansatz, thereby removing contributions 
of statistical and multiplicity fluctuations, as well as detector effects.
\color{black}
\fi
Large relative fluctuations of about \mbox{$40$--$50$\%} are found,
independent of centrality. The new data are well described by fluctuations 
in the shape of the initial collision region, as predicted with the participant 
eccentricity using MCG simulations. These results substantiate our conclusions 
from previous studies on the relevance of such fluctuations for the elliptic 
flow across nuclear species: The initial-geometry seems to drive the hydrodynamic 
evolution of the system, not only on average, but event-by-event.

\ifshort
\else
\color{\mycol}
\section*{Acknowledgments}
%
%
%
%
This work was partially supported by U.S. DOE grants 
DE-AC02-98CH10886,
DE-FG02-93ER40802, 
DE-FG02-94ER40818,  
DE-FG02-94ER40865, 
DE-FG02-99ER41099, and
DE-AC02-06CH11357, by U.S. 
NSF grants 9603486, 
0072204,            
and 0245011,        
by Polish KBN grant 1-P03B-062-27(2004-2007),
by NSC of Taiwan Contract NSC 89-2112-M-008-024, and
by Hungarian OTKA grant (F 049823).
\color{black}
\fi

\end{document}